\documentclass[11pt,preprintnumbers,titlepage,nofootinbib]{revtex4-2}
\usepackage[latin9]{inputenc}
\usepackage{amsmath}
\usepackage{amssymb}
\usepackage{graphicx}
\usepackage[unicode=true,
 bookmarks=false,
 breaklinks=false,pdfborder={0 0 1},backref=section,colorlinks=false]
 {hyperref}
\hypersetup{
 colorlinks,linkcolor=blue,citecolor=blue,urlcolor=blue}

\makeatletter

\usepackage{wasysym}
\usepackage{array}
\usepackage{multirow}
\usepackage{wasysym}
\usepackage{wasysym}
\usepackage{amsfonts}
\usepackage{subfigure}
\usepackage{cleveref}
\usepackage{listings}
\usepackage{xcolor}
\usepackage{array}
\usepackage{url}
\usepackage{color}

\@ifundefined{textcolor}{}{
\definecolor{BLACK}{gray}{0}
\definecolor{WHITE}{gray}{1}
\definecolor{RED}{rgb}{1,0,0}
\definecolor{GREEN}{rgb}{0,1,0}
\definecolor{BLUE}{rgb}{0,0,1}
\definecolor{CYAN}{cmyk}{1,0,0,0}
\definecolor{MAGENTA}{cmyk}{0,1,0,0}
\definecolor{YELLOW}{cmyk}{0,0,1,0}
}

\makeatother

\begin{document}
\preprint{CTP-SCU/2024002}
\title{Nonlinear Stability of Black Holes with a Stable Light Ring}
\author{Guangzhou Guo$^{a}$}
\email{guogz@sustech.edu.cn}

\author{Peng Wang$^{b}$}
\email{pengw@scu.edu.cn}

\author{Yupeng Zhang$^{c}$}
\email{zhangyupeng@lzu.edu.cn}

\affiliation{$^{a}$Department of Physics, Southern University of Science and Technology,
Shenzhen, 518055, China}
\affiliation{$^{b}$Center for Theoretical Physics, College of Physics, Sichuan
University, Chengdu, 610064, China}
\affiliation{$^{c}$Key Laboratory of Quantum Theory and Applications of MoE, Lanzhou
Center for Theoretical Physics, Key Laboratory of Theoretical Physics
of Gansu Province, Institute of Theoretical Physics $\&$ Research
Center of Gravitation, Lanzhou University, Lanzhou, 730000, China}
\begin{abstract}
Recently, ultracompact objects have been found to be susceptible to
a new nonlinear instability, known as the light-ring instability,
triggered by stable light rings. This discovery raises concerns about
the viability of these objects as alternatives to black holes. In
this work, we investigate the presence of the light-ring instability
in scalarized Reissner-Nordström black holes, which have been previously
shown to admit stable light rings. We employ fully nonlinear numerical
evolutions of both scalarized black holes with and without stable
light rings, perturbing them initially with spherically symmetric
scalar perturbations. Our simulations demonstrate the long-term stability
of these scalarized black holes, suggesting that the presence of a
stable light ring may not necessarily induce the light-ring instability.
\end{abstract}
\maketitle
\tableofcontents{}

{}

{}

{}

\section{Introduction}

The past decade has witnessed remarkable progress in the field of
black hole physics, driven by the groundbreaking detection of gravitational
waves from binary black hole mergers \cite{Abbott:2016blz}. This
discovery has opened unprecedented avenues for exploring the intricacies
of black holes, particularly through the analysis of quasinormal modes
during the ringdown phase, providing valuable insights into the properties
of black hole spacetime \cite{Nollert:1999ji,Berti:2007dg,Cardoso:2016rao,Price:2017cjr,Giesler:2019uxc}.
Furthermore, the Event Horizon Telescope collaboration has revolutionized
our understanding of black holes by capturing the first images of
M87{*} and Sgr A{*}, revealing a striking feature: a luminous ring
encircling a dark shadow \cite{Akiyama:2019cqa,Akiyama:2019brx,Akiyama:2019sww,Akiyama:2019bqs,Akiyama:2019fyp,Akiyama:2019eap,Akiyama:2021qum,Akiyama:2021tfw,EventHorizonTelescope:2022xnr,EventHorizonTelescope:2022vjs,EventHorizonTelescope:2022wok,EventHorizonTelescope:2022exc,EventHorizonTelescope:2022urf,EventHorizonTelescope:2022xqj}.
These distinctive signatures have been attributed to the intense light
deflection occurring near unstable bound photon orbits, known as light
rings. Moreover, recent studies have established a strong connection
between light rings and a specific class of quasinormal modes of perturbations
in the black hole spacetime \cite{Ferrari:1984zz,Cardoso:2008bp,Yang:2012he,Konoplya:2017wot,Jusufi:2019ltj,Cuadros-Melgar:2020kqn,Qian:2021aju}.

While current observations largely agree with general relativity's
predictions, limitations in detection resolution motivate investigations
into alternative theories of gravity. In particular, Exotic Compact
Objects (ECOs) have attracted interest due to their ability to mimic
black holes \cite{Lemos:2008cv,Cunha:2017wao,Cunha:2018acu,Shaikh:2018oul,Dai:2019mse,Huang:2019arj,Simonetti:2020ivl,Wielgus:2020uqz,Yang:2021diz,Bambi:2021qfo,Peng:2021osd,Ghosh:2022mka,AbhishekChowdhuri:2023ekr}.
These objects harbor stable light rings, which trap specific quasinormal
modes with exceptionally long lifetimes and vanishingly small imaginary
components \cite{Cardoso:2014sna}. The reflective nature of ECO boundaries
leads to the generation of echo signals during the post-merger ringdown
phase of binary black hole mergers, with these echoes being dominated
by the long-lived modes mentioned above \cite{Bueno:2017hyj,Mark:2017dnq,Rahman:2021kwb}.
Notably, recent LIGO/Virgo data hints at the presence of such echoes
in gravitational wave signals from binary black hole mergers \cite{Abedi:2016hgu,Abedi:2017isz}.
This evidence, while intriguing, requires further investigation to
confirm the existence of ECOs and their associated echo signals.

Indeed, the viability of ECOs with stable light rings remains under
scrutiny due to concerns about instabilities arising from both linear
and nonlinear effects \cite{Holzegel:2011uu,Cardoso:2014sna,Keir:2014oka,Cunha:2022gde}.
In rotating ECOs, the presence of an ergoregion can trigger linear
ergoregion instabilities, leading to the amplification of long-lived
quasinormal modes \cite{Cardoso:2014sna}. Even within dissipative
systems where linear perturbations are expected to decay, stable light
rings can trap these modes such that their decay is slower than logarithmic
\cite{Keir:2014oka}. These long-lived perturbations residing near
stable light rings could trigger novel nonlinear instabilities, known
as light-ring instability \cite{Keir:2014oka,Cunha:2022gde}. Recent
studies employing fully nonlinear numerical simulations in parameter
spaces free from linear ergoregion instabilities have conclusively
demonstrated the existence of the light-ring instability. These instabilities
drive ECOs to either migrate towards configurations lacking stable
light rings or collapse into black holes \cite{Cunha:2022gde}.

Meanwhile, to understand the formation of hairy black holes, researchers
have explored a class of Einstein-Maxwell-scalar (EMS) models \cite{Herdeiro:2018wub}.
These models incorporate non-minimal couplings between the scalar
and electromagnetic fields, leading to instabilities that can trigger
the spontaneous growth of a ``hair''\ -- a scalar field configuration
around the black hole. Using fully nonlinear numerical simulations,
Herdeiro et al. demonstrated the transformation of Reissner-Nordström
(RN) black holes into scalarized RN black holes \cite{Herdeiro:2018wub}.
This discovery has ignited a surge of research within the EMS framework,
exploring diverse aspects such as different non-minimal coupling functions
\cite{Fernandes:2019rez,Fernandes:2019kmh,Blazquez-Salcedo:2020nhs},
massive and self-interacting scalar fields \cite{Zou:2019bpt,Fernandes:2020gay},
horizonless reflecting stars \cite{Peng:2019cmm}, stability analysis
of scalarized black holes \cite{Myung:2018vug,Myung:2019oua,Zou:2020zxq,Myung:2020etf,Mai:2020sac},
higher dimensional scalar-tensor models \cite{Astefanesei:2020qxk},
quasinormal modes of scalarized black holes \cite{Myung:2018jvi,Blazquez-Salcedo:2020jee},
two U(1) fields \cite{Myung:2020dqt}, quasitopological electromagnetism
\cite{Myung:2020ctt}, topology and spacetime structure influences
\cite{Guo:2020zqm}, scalarized black hole solutions in the dS/AdS
spacetime \cite{Brihaye:2019dck,Brihaye:2019gla,Zhang:2021etr,Guo:2021zed,Chen:2023eru},
dynamical scalarization and descalarization \cite{Zhang:2021nnn,Zhang:2022cmu,Jiang:2023yyn}
and rotating scalarized black hole solutions \cite{Guo:2023mda}.

Intriguingly, within specific parameter ranges, scalarized RN black
holes can possess two unstable light rings and one stable light ring
on the equatorial plane outside their event horizons \cite{Gan:2021pwu}.
This unique feature has spurred investigations into the optical signatures
of various phenomena near these black holes, including accretion disks
\cite{Gan:2021pwu,Gan:2021xdl,Chen:2023qic}, luminous celestial spheres
\cite{Guo:2022muy}, infalling stars \cite{Chen:2022qrw} and hot
spots \cite{Chen:2024ilc}. Studies have shown that the presence of
an additional unstable light ring can significantly increase the observed
flux from accretion disks, create beat signals in the visibility amplitude,
generate triple higher-order images of luminous celestial spheres,
and trigger a cascade of additional flashes from an infalling star.
However, the existence of a stable light ring raises concerns about
spacetime stability due to the potential presence of long-lived quasinormal
modes \cite{Guo:2021bcw,Guo:2021enm,Guo:2022umh,Zhong:2022jke}. Recent
work has demonstrated that the stable light ring can give rise to
superradiance instabilities associated with charged scalar perturbations
\cite{Guo:2023ivz}. Moreover, the existence of mutiple light rings
has also been found in other black hole scenarios, including dyonic
black holes with a quasi-topological electromagnetic term \cite{Liu:2019rib,Huang:2021qwe},
black holes in massive gravity \cite{deRham:2010kj,Dong:2020odp}
and wormholes in the black-bounce spacetime \cite{Tsukamoto:2021caq,Tsukamoto:2021fpp,Tsukamoto:2022vkt}.
For a comprehensive analysis of black holes with multiple light rings,
we refer readers to \cite{Guo:2022ghl}.

This paper investigates the nonlinear stability of scalarized RN black
holes, aiming to elucidate the fate of stable light rings within black
hole spacetimes. The paper is structured as follows. In Section \ref{sec:SETUP},
we introduces the EMS model, including the construction of static
scalarized black hole solutions and their dynamic evolution. We present
numerical results for the spontaneous scalarization of RN black holes
and dynamic stability analysis of scalarized RN black holes in Section
\ref{sec:NS}. Finally, Section \ref{Sec:Conc} presents our conclusions.
Throughout this paper, we adopt the convention $G=c=4\pi\epsilon_{0}=1$.

\section{Set Up}

\label{sec:SETUP}

This section begins with a brief overview of the EMS model, where
a tachyonic instability can trigger the spontaneous scalarization
of RN black holes. To investigate this phenomenon, we construct static
scalarized black hole solutions within the EMS framework. We then
derive the effective potentials governing both photon and scalar field
perturbations. Finally, we establish the full nonlinear dynamics within
the EMS model, providing a framework to investigate the evolution
of both RN and scalarized black holes.

\subsection{The EMS Model}

The EMS model incorporates a non-minimal coupling function between
the scalar and electromagnetic fields, denoted by $f\left(\phi\right)$.
This coupling can induce tachyonic instabilities, leading to the spontaneous
formation of scalarized black holes. We explore this phenomenon within
the framework of the EMS action, 
\begin{equation}
S=\frac{1}{16\pi}\int d^{4}x\sqrt{-g}\left[R-2\partial_{\mu}\phi\partial^{\mu}\phi-f\left(\phi\right)F^{\mu\nu}F_{\mu\nu}\right],\label{eq:Action}
\end{equation}
where $F_{\mu\nu}=\partial_{\mu}A_{\nu}-\partial_{\nu}A_{\mu}$ denotes
the electromagnetic field strength tensor, and $f\left(\phi\right)=e^{\alpha\phi^{2}}$.

In a scalar-free background (i.e., RN black holes), a scalar perturbation
$\delta\phi$ follows the linearized equation of motion,
\begin{equation}
\left(\square-\mu_{\text{eff}}^{2}\right)\delta\phi=0,\label{eq:delta phi}
\end{equation}
where the effective mass square $\mu_{\text{eff}}^{2}=-\alpha Q^{2}/r^{4}$,
and $Q$ represents the RN black hole charge. Notably, a positive
coupling constant $\alpha$ leads to a negative effective mass squared
$\mu_{\text{eff}}^{2}$, potentially triggering tachyonic instabilities
for the scalar field in RN black hole. These instabilities, as demonstrated
in \cite{Herdeiro:2018wub,Guo:2021zed}, can initiate spontaneous
scalarization, transforming the RN black holes into scalarized ones.

We obtain the equations of motion by varying the action $\left(\ref{eq:Action}\right)$
with respect to the metric field $g_{\mu\nu}$, the scalar field $\phi$
and the electromagnetic field $A_{\mu}$, 
\begin{align}
R_{\mu\nu}-\frac{1}{2}Rg_{\mu\nu} & =2T_{\mu\nu},\nonumber \\
\square\phi-\frac{\alpha}{2}\phi e^{\alpha\phi^{2}}F^{\mu\nu}F_{\mu\nu} & =0,\label{eq:nonlinearEOMs}\\
\partial_{\mu}\left(\sqrt{-g}e^{\alpha\phi^{2}}F^{\mu\nu}\right) & =0,\nonumber 
\end{align}
where the energy-momentum tensor $T_{\mu\nu}$ is given by 
\begin{equation}
T_{\mu\nu}=\partial_{\mu}\phi\partial_{\nu}\phi-\frac{1}{2}g_{\mu\nu}\left(\partial\phi\right)^{2}+e^{\alpha\phi^{2}}\left(F_{\mu\rho}F_{\nu}^{\text{ }\rho}-\frac{1}{4}g_{\mu\nu}F_{\rho\sigma}F^{\rho\sigma}\right).
\end{equation}

\subsection{Static Scalarized Black Holes}

\label{subsec:Static-Black-Hole}

\subsubsection{Black Hole Solution}

To construct the static scalarized black hole solution, we consider
the asymptotically flat and spherically symmetric ansatz \cite{Herdeiro:2018wub,Guo:2021zed},
\begin{align}
ds^{2} & =-N(r)e^{-2\delta(r)}dt^{2}+\frac{1}{N(r)}dr^{2}+r^{2}\left(d\theta^{2}+\sin^{2}\theta d\varphi^{2}\right),\nonumber \\
A_{\mu}dx^{\mu} & =V(r)dt\text{ and}\ \phi=\phi_{s}(r).\label{eq:sHBH}
\end{align}
Substituting the ansatz $\left(\ref{eq:sHBH}\right)$ into the equations
of motion $\left(\ref{eq:nonlinearEOMs}\right)$, one obtains 
\begin{align}
N^{\prime}(r) & =\frac{1-N(r)}{r}-\frac{Q^{2}}{r^{3}e^{\alpha\phi_{s}^{2}(r)}}-rN(r)\left[\phi_{s}^{\prime}(r)\right]^{2},\nonumber \\
\left[r^{2}N(r)\phi_{s}^{\prime}(r)\right]^{\prime} & =-\frac{\alpha Q^{2}\phi_{s}(r)}{r^{2}e^{\alpha\phi_{s}^{2}(r)}}-r^{3}N(r)\left[\phi_{s}^{\prime}(r)\right]^{3},\nonumber \\
\delta^{\prime}(r) & =-r\left[\phi_{s}^{\prime}(r)\right]^{2},\label{eq:sEOM}\\
V^{\prime}(r) & =\frac{Q}{r^{2}e^{\alpha\phi_{s}^{2}(r)}}e^{-\delta(r)},\nonumber 
\end{align}
where primes denote derivatives with respect to $r$, and the integration
constant $Q$ represents the black hole charge. To solve for static
black hole solutions from eqn. $\left(\ref{eq:sEOM}\right)$, one
needs to impose appropriate boundary conditions on the event horizon
and spatial infinity. On the event horizon $r_{h}$, the black hole
solution is characterized by 
\begin{equation}
N(r_{h})=0\text{, }\delta(r_{h})=\delta_{0}\text{, }\phi_{s}(r_{h})=\phi_{0}\text{, }V(r_{h})=V_{0}\text{,}\label{eq:rh condition}
\end{equation}
where $V_{0}$ is the electrostatic potential. At spatial infinity,
the black hole solution has asymptotic behaviors, 
\begin{equation}
N(r)=1-\frac{2M}{r}+...\text{,}\ \delta(r)=\frac{Q_{s}^{2}}{2r^{2}}+...\text{, }\phi_{s}(r)=\frac{Q_{s}}{r}+...\text{, }V(r)=-\frac{Q}{r}+...\text{,}\label{eq:infinity condition}
\end{equation}
where $M$ is the black hole mass, and $Q_{s}$ denotes the scalar
charge. 
\begin{figure}[t]
\begin{centering}
\includegraphics[scale=0.76]{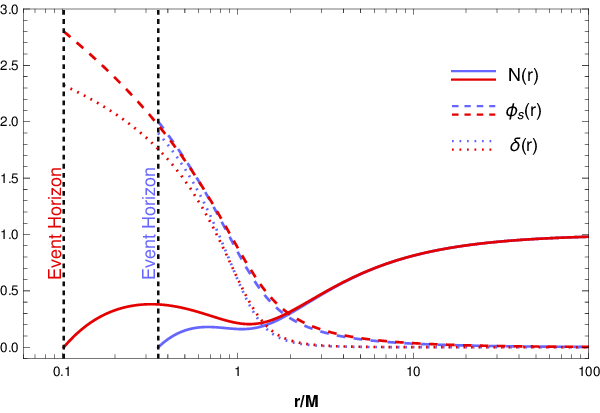} \includegraphics[scale=0.8]{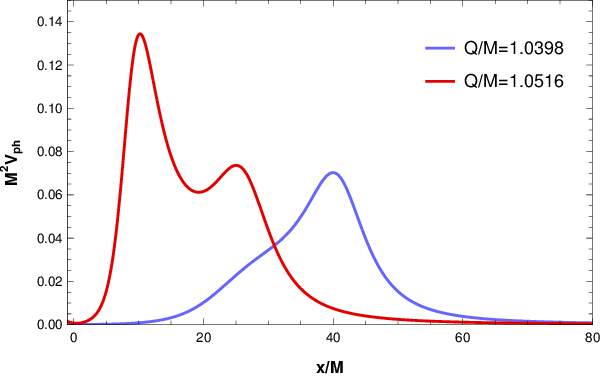} 
\par\end{centering}
\caption{Static scalarized black holes with $\alpha=0.8$ are shown for $Q/M=1.0398$
(blue line) and $Q/M=1.0526$ (red line). \textbf{Left Panel}: Metric
functions $N\left(r\right)$, $\phi_{s}\left(r\right)$ and $\delta\left(r\right)$
are plotted outside the event horizon (dashed line). \textbf{Right
Panel}: The effective potential of photons versus the tortoise coordinate
$x$ exhibits a double-peak structure for $Q/M=1.0526$, indicating
a stable light ring on the equatorial plane.}
\label{bhsols}
\end{figure}

This paper employs the shooting method to solve $\left(\ref{eq:sEOM}\right)$
for static black hole solutions that fulfill the boundary conditions
outlined in eqns. $\left(\ref{eq:rh condition}\right)$ and $\left(\ref{eq:infinity condition}\right)$.
Notably, the equations of motion $\left(\ref{eq:sEOM}\right)$ allow
for a scalar-free solution with $\phi_{0}=\delta_{0}=0$, corresponding
to RN black holes with $\phi=0$. Moreover, solutions with a non-trivial
scalar field ($\phi\neq0$) can also be obtained, resulting in hairy
black holes characterized by non-zero values of $\phi_{0}$ and $\delta$.
The left panel of Fig. $\ref{bhsols}$ depicts the metric functions
of two such static scalarized black holes with a coupling constant
$\alpha=0.8$. The blue and red lines represent solutions for $Q/M=1.0398$
and $Q/M=1.0526$, respectively.

\subsubsection{Light Rings}

To analyze the trajectory of photons around a static and spherically
symmetric black hole, it suffices to focus on the equatorial plane
with $\theta=\pi/2$. To identify null circular geodesics (light rings)
on this plane, we begin with the Lagrangian of a photon \cite{Guo:2022muy},
\begin{equation}
\mathcal{L}=\frac{1}{2}\left(-N\left(r\right)e^{-2\delta\left(r\right)}\dot{t}^{2}+\frac{1}{N\left(r\right)}\dot{r}^{2}+r^{2}\dot{\varphi}^{2}\right),\label{eq:photon Lag}
\end{equation}
where dots denote time derivatives with respect to the affine parameter
$\tau$, and $\mathcal{L}=0$ describes the photon's motion. Since
the metric $\left(\ref{eq:sHBH}\right)$ is independent of $t$ and
$\varphi$, the black hole spacetime possesses two Killing vectors
$\partial_{t}$ and $\partial_{\varphi}$, leading to conserved energy
$E$ and angular momentum $L$, respectively. These conserved quantities
are derived from the photon's generalized canonical momenta via the
Lagrangian $\left(\ref{eq:photon Lag}\right)$, 
\begin{align}
E & =-p_{t}=N\left(r\right)e^{-2\delta\left(r\right)}\dot{t},\nonumber \\
L & =p_{\varphi}=r^{2}\dot{\varphi}.\label{eq:E and L}
\end{align}
Using $\mathcal{L}=0$, the radial equation of motion for the photon
can be expressed in term of $E$ and $L$, 
\begin{equation}
e^{-2\delta\left(r\right)}\dot{r}^{2}=E^{2}-V_{\text{ph}}\left(r\right)L^{2},
\end{equation}
where $V_{\text{ph}}\left(r\right)$ is the effective potential of
photons defined as 
\begin{equation}
V_{\text{ph}}\left(r\right)=\frac{e^{-2\delta\left(r\right)}N\left(r\right)}{r^{2}}.\label{eq:Vph}
\end{equation}
Therefore, a light ring with radius $r_{c}$ exists where $V_{\text{ph}}\left(r_{c}\right)=E^{2}/L^{2}$
and $V_{\text{ph}}^{\prime}\left(r_{c}\right)=0$. Furthermore, maxima
and minima of $V_{\text{eff}}(r)$ correspond to unstable and stable
light rings, respectively.

The right panel of Fig. $\ref{bhsols}$ depicts the effective photon
potential $V_{\text{ph}}\left(r\right)$ for the static scalarized
black holes with $\alpha=0.8$ (blue line: $Q/M=1.0398$; red line:
$Q/M=1.0526$). For the lower black hole charge (blue line), $V_{\text{ph}}\left(r\right)$
exhibits a single maximum, indicating an unstable light ring on the
equatorial plane. Interestingly, the higher charge (red line) results
in two maxima and a minimum, corresponding to two unstable light rings
and one stable light ring on the equatorial plane. Due to spherical
symmetry, these light rings translate to unstable and stable photon
spheres around the black holes.

\subsubsection{Scalar Perturbations}

We investigate the linear stability of static scalarized black holes
by analyzing spherically symmetric perturbations in the black hole
spacetime \cite{Guo:2021zed}. The adopted ansatz incorporating time-dependent
perturbations is given by 
\begin{align}
ds^{2} & =-\tilde{N}(t,r)e^{-2\tilde{\delta}(t,r)}dt^{2}+\frac{1}{\tilde{N}(t,r)}dr^{2}+r^{2}\left(d\theta^{2}+\sin^{2}\theta d\varphi^{2}\right),\nonumber \\
A_{\mu}dx^{\mu} & =\tilde{V}(t,r)dt\text{ and}\ \phi=\tilde{\phi}(t,r),\label{eq:pHBH}
\end{align}
where the metric functions, electromagnetic field and scalar field
are separated as 
\begin{align}
\tilde{N}(t,r) & =N(r)+\epsilon N_{1}\left(t,r\right),\quad\tilde{\delta}(t,r)=\delta(r)+\epsilon\delta_{1}(t,r),\nonumber \\
\tilde{V}(t,r) & =V(r)+\epsilon V_{1}(t,r),\quad\tilde{\phi}(t,r)=\phi_{s}(r)+\epsilon\phi_{1}(t,r).\label{eq:pers}
\end{align}
Solving eqn. $\left(\ref{eq:nonlinearEOMs}\right)$ with the ansatz
$\left(\ref{eq:pHBH}\right)$ yields the linearized equation of motion
for the scalar perturbation in the time domain, 
\begin{equation}
\left(-\frac{\partial^{2}}{\partial t^{2}}+\frac{\partial^{2}}{\partial x^{2}}-V_{\text{sc}}\left(r\right)\right)\Psi\left(t,r\right)=0,\label{eq:eom_pert_t}
\end{equation}
where $\Psi(t,r)=r\phi_{1}(t,r)$, and the tortoise coordinate $x$
is defined by $dx/dr=e^{\delta(r)}/N(r)$. The effective potential
for the scalar perturbation is given by 
\begin{equation}
V_{\text{sc}}=\frac{e^{-2\delta}N}{r^{2}}\left[1-N-2r^{2}\phi_{s}^{\prime2}-\frac{Q^{2}}{r^{2}e^{\alpha\phi_{s}^{2}}}\left(1+\alpha-2r^{2}\phi_{s}^{\prime2}+4\alpha r\phi_{s}\phi_{s}^{\prime}-2\alpha^{2}\phi_{s}^{2}\right)\right].\label{eq:Vsc}
\end{equation}
It is important to note that this analysis only considers spherical
perturbations, leading to the decoupling of the scalar perturbation
from the gravitational and electromagnetic ones. However, including
non-spherical perturbations would introduce coupling between the scalar
and other types of perturbations \cite{Myung:2019oua}. To numerically
solve the partial differential equation $\left(\ref{eq:eom_pert_t}\right)$
for the evolution of the linear perturbation $\Psi$, we employ a
small Gaussian perturbation as the initial condition and utilize radiative
boundary conditions.

To determine the frequency $\omega$ of quasinormal modes for the
scalar perturbation, we perform a Fourier transformation $\Psi(t,x)=\int d\omega\hat{\Psi}(\omega,x)e^{-i\omega t}$.
Consequently, eqn. $\left(\ref{eq:eom_pert_t}\right)$ transforms
into the equation for $\hat{\Psi}$ in the frequency domain, 
\begin{equation}
\left(\frac{\partial^{2}}{\partial x^{2}}+\omega^{2}-V_{\text{sc}}\left(r\right)\right)\hat{\Psi}\left(\omega,x\right)=0.\label{eq:eom_pert_omega}
\end{equation}
Imposing ingoing and outgoing boundary conditions at the event horizon
and spatial infinity, respectively, 
\begin{align}
\hat{\Psi}\left(\omega,x\right) & \sim e^{-i\omega x},\qquad x\rightarrow-\infty,\nonumber \\
\hat{\Psi}\left(\omega,x\right) & \sim e^{i\omega x},\qquad x\rightarrow+\infty,\label{eq:in-out bds}
\end{align}
we obtain a discrete set of quasinormal modes with non-vanishing imaginary
parts. These imaginary parts indicate the linear stability of the
system: a negative value signifies a dissipative and stable system,
while a positive value signifies an unstable mode. In this work, we
numerically solve eqn. $\left(\ref{eq:eom_pert_omega}\right)$ for
quasinormal modes using direct integration. It is noteworthy that
the scalar effective potential $\left(\ref{eq:Vsc}\right)$ reduces
to the photon effective potential $\left(\ref{eq:Vph}\right)$ in
the eikonal limit, except for a prefactor \cite{Guo:2022ghl}. In
this limit, studies have found long-lived modes with an exponentially
small imaginary part residing near stable light rings, potentially
leading to specific types of nonlinear instabilities \cite{Holzegel:2011uu,Cardoso:2014sna,Keir:2014oka,Cunha:2022gde}.

\subsection{Numerical Evolutions}

Analyzing the time evolution of black holes in spherically symmetric
spacetimes is facilitated by employing Painlevé-Gullstrand-like (PG)
coordinates. These coordinates utilize a time-dependent ansatz, as
given by \cite{Zhang:2021nnn,Jiang:2023yyn}, 
\begin{align}
ds^{2} & =-\left[1-\zeta^{2}(t,r)\right]\beta^{2}(t,r)dt^{2}+2\zeta(t,r)\beta(t,r)dtdr+dr^{2}+r^{2}\left(d\theta^{2}+\sin^{2}\theta d\varphi^{2}\right),\nonumber \\
A_{\mu}dx^{\mu} & =A(t,r)dt\text{ and}\ \phi=\phi(t,r).\label{eq:dansatz}
\end{align}
When the black hole reaches equilibrium, the dynamic metric $\left(\ref{eq:dansatz}\right)$
becomes time-independent and reduces to the aforementioned static
metric $\left(\ref{eq:sHBH}\right)$ through a coordinate transformation
\cite{Jiang:2023yyn}, 
\begin{equation}
\left.dt\right\vert _{\text{spherical coordinates}}\rightarrow\left.dt-\frac{\zeta}{\left(1-\zeta^{2}\right)\beta}dr\right\vert _{\text{PG coordinates}}.
\end{equation}
This transformation leads to the following relationship between the
metric functions, 
\begin{equation}
N=1-\zeta^{2},\;e^{-\delta}=\beta.\label{eq:DtoS}
\end{equation}
To denote the apparent horizon during the black hole's evolution,
we use $r_{h}$, which is identified at each time slice by solving
the equation $\zeta(t,r_{h})=1$. For numerical stability purposes,
an auxiliary variable for the scalar field is introduced \cite{Zhang:2021nnn},
\begin{equation}
\Pi\left(t,r\right)=\frac{1}{\beta\left(t,r\right)}\partial_{t}\phi\left(t,r\right)-\zeta\left(t,r\right)\phi^{\prime}(t,r).\label{eq:Pi}
\end{equation}

Substituting eqns. $\left(\ref{eq:dansatz}\right)$ and $\left(\ref{eq:Pi}\right)$
into eqn. $\left(\ref{eq:nonlinearEOMs}\right)$, the equations of
motion for the gravitational and scalar fields become 
\begin{align}
\zeta^{\prime} & =\frac{r}{2\zeta}\left(\phi^{\prime2}+\Pi^{2}\right)+\frac{Q^{2}}{2r^{3}\zeta e^{\alpha\phi^{2}}}+r\Pi\phi^{\prime}-\frac{\zeta}{2r},\nonumber \\
\beta^{\prime} & =-\frac{r\Pi\phi^{\prime}\beta}{\zeta},\label{eq:metric eqs}\\
\partial_{t}\zeta & =\frac{r\beta}{\zeta}\left(\Pi+\phi^{\prime}\zeta\right)\left(\Pi\zeta+\phi^{\prime}\right),\nonumber 
\end{align}
and 
\begin{align}
\partial_{t}\phi & =\beta\left(\Pi+\phi^{\prime}\zeta\right),\nonumber \\
\partial_{t}\Pi & =\frac{\left[\left(\Pi\zeta+\phi^{\prime}\right)\beta r^{2}\right]^{\prime}}{r^{2}}+\frac{\alpha Q^{2}}{r^{4}e^{\alpha\phi^{2}}}\phi\beta,\label{eq:scalar eqs}
\end{align}
respectively. The equation of motion for the electromagnetic field
then becomes 
\begin{equation}
A^{\prime}=\frac{Q\beta}{r^{2}e^{\alpha\phi^{2}}},\label{eq:A eq}
\end{equation}
indicating that $A(t,r)$ can be determined by eqn. $\left(\ref{eq:A eq}\right)$
once the metric functions and the scalar field are obtained.\ The
evolution of the EMS system is achieved by numerically integrating
the last equation in eqn. $\left(\ref{eq:metric eqs}\right)$ and
both equations in eqn. $\left(\ref{eq:scalar eqs}\right)$ using the
fourth-order Runge-Kutta method. After obtaining $\zeta$, $\phi$
and $\Pi$ at each timestep, the second equation of eqn. $\left(\ref{eq:metric eqs}\right)$
is solved for the lapse function $\beta$, enforcing the boundary
condition $\left.\beta\right\vert _{r\rightarrow\infty}=1$. The first
equation of eqn. $\left(\ref{eq:metric eqs}\right)$ serves as a constraint
equation, allowing for the assessment of numerical simulation errors.

Our numerical scheme extends the evolution domain beyond the event
horizon of the initial black hole, encompassing a small interior region.
This choice ensures that the domain always covers the exterior region
throughout the evolution, as the apparent horizon $r_{h}$ never shrinks.
The domain is then truncated at a sufficiently distant region where
radiative boundary conditions are applied to the evolved fields, $\zeta$,
$\phi$ and $\Pi$. Both RN and scalarized black holes are considered
as initial states in our numerical computations. We introduce a Gaussian
perturbation $\delta\phi$ to the scalar field of the initial state,
which serves as the initial data for the scalar field. This perturbation
is described by 
\begin{equation}
\delta\phi=pe^{-\frac{\left(r-r_{0}\right)^{2}}{\Delta^{2}}},\label{eq:initial perts}
\end{equation}
where $p$, $r_{0}$ and $\Delta$ represent the perturbation's amplitude,
location and width, respectively. The initial metric functions $\zeta$
and $\beta$ are then determined by solving eqn. $\left(\ref{eq:metric eqs}\right)$
with the given scalar field initial data and the initial state's $\Pi$.
In this study, we have integrated the numerical computations of eqns.
$\left(\ref{eq:metric eqs}\right)$ and $\left(\ref{eq:scalar eqs}\right)$
into the Einstein Toolkit, a framework renowned for its efficiency
in simulating black hole evolutions \cite{Zachariah}.

\section{Numerical Simulation}

\label{sec:NS}

This section first investigates the spontaneous scalarization from
RN black holes, elucidating the formation process of scalarized black
holes. Notably, the evolution from RN black holes to scalarized black
holes has been previously explored in \cite{Herdeiro:2018wub}, providing
a benchmark for validating our numerical computations. We then perform
fully nonlinear numerical evolutions starting from both scalarized
black holes with and without a stable light ring to assess their stability
under spherical perturbations.

For a scalar field characterized by an unstable mode with frequency
$\omega$, the initial growth of the field at the event horizon, denoted
by $\phi_{h}$, can be approximated as 
\begin{equation}
\phi_{h}\left(t\right)\approx\phi_{h}\left(t_{0}\right)+h\left(p\right)e^{-i\omega\left(t-t_{0}\right)},
\end{equation}
where $h\left(p\right)$ depends on the amplitude $p$ of the initial
Gaussian perturbation $\left(\ref{eq:initial perts}\right)$ \cite{Zhang:2021nnn,Zhang:2022cmu}.
Studies have shown that a larger $p$ leads to a faster growth stage
of the scalar field \cite{Zhang:2021nnn,Chen:2023iws}. To gain a
deeper understanding of the black hole evolution, we calculate the
quantity $\ln\left\vert d\phi_{h}/dt\right\vert $ during the numerical
simulation. The imaginary part of the unstable mode $\omega$ can
then be identified by matching it with the slope of $\ln\left\vert d\phi_{h}/dt\right\vert $
during the scalar field growth stage \cite{Chen:2023iws}. Additionally,
consistent with the second law of thermodynamics, the area of the
black hole's apparent horizon $A_{h}=4\pi r_{h}^{2}$ never decreases
during the simulation, serving as a further check on our numerical
results.

\subsection{Spontaneous Scalarization}

\begin{figure}[ptb]
\begin{centering}
\includegraphics[scale=0.77]{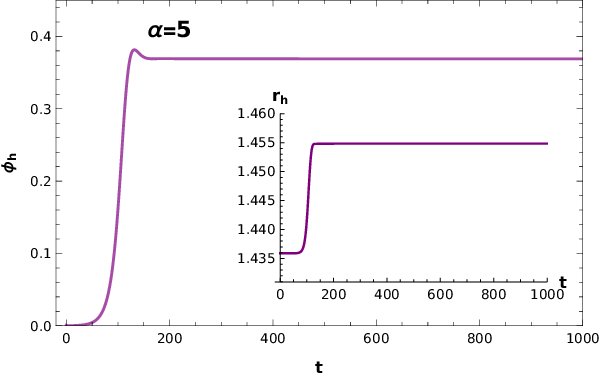} \includegraphics[scale=0.77]{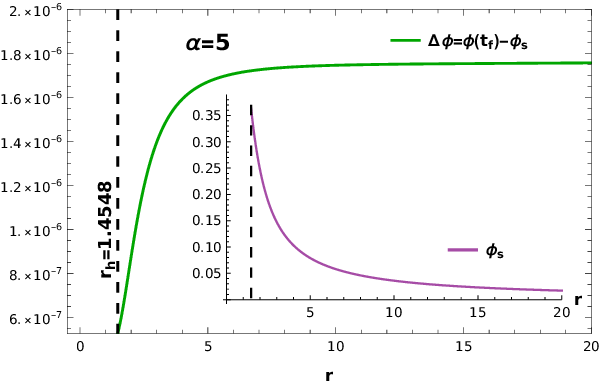} 
\par\end{centering}
\begin{centering}
\includegraphics[scale=0.77]{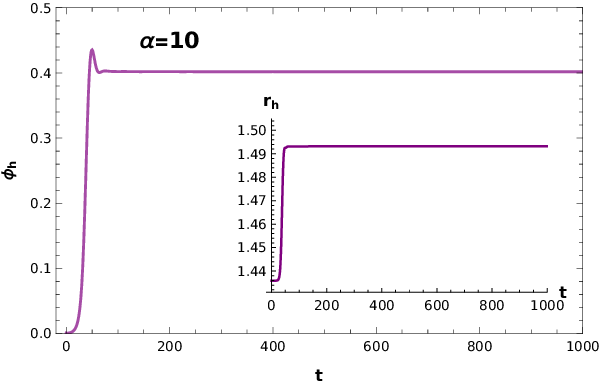} \includegraphics[scale=0.77]{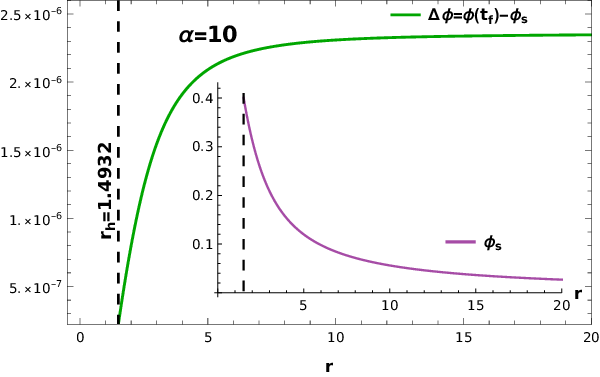} 
\par\end{centering}
\caption{Spontaneous scalarization in the EMS model for coupling constants
$\alpha=5$ (\textbf{Upper Row}) and $\alpha=10$ (\textbf{Lower}
\textbf{Row}). A RN black hole with $Q=0.9$ and $M=1$ serves as
the initial state. \textbf{Left Column}: The scalar field value $\phi_{h}$
at the apparent horizon $r_{h}$ exhibits growth due to the tachyonic
instability. Notably, a larger coupling constant $\alpha$ (e.g.,
$\alpha=10$ in the lower panel) leads to a more pronounced growth,
accelerating the black hole's scalarization. The apparent horizon
$r_{h}$ consistently increases throughout the simulation, complying
with the second law of thermodynamics. \textbf{Right Column}: The
main plots depict the difference $\triangle\phi$ between the scalar
field of the end state, $\phi\left(t_{f}\right)$ at $t_{f}=1000$,
and that of a corresponding static scalarized black hole, $\phi_{s}$,
shown in the insets. The small magnitude of $\triangle\phi$ (around
$10^{-6}$ or below) validates the accuracy of our numerical results
and suggests that the final equilibrium states closely resemble static
scalarized black holes. The vertical black dashed lines represent
the horizons.}
\label{RNBH}
\end{figure}

Fig. \ref{RNBH} presents the spontaneous scalarization process for
an initial RN black hole with $Q=0.9$ and $M=1$. The upper and lower
rows depict cases with $\alpha=5$ and $\alpha=10$, respectively.
The left column shows the dynamical evolution of the scalar field
at the apparent horizon $\phi_{h}$ alongside the evolution of the
apparent horizon radius $r_{h}$ in the inset of each panel. Consistent
with the second law of thermodynamics, $r_{h}$ never decreases throughout
the simulation. The plots reveal initial scalar field growth due to
tachyonic instabilities, followed by stabilization at an equilibrium
state via nonlinear effects. To characterize the equilibrium state,
we consider a static scalarized black hole with matching horizon radius
and scalar field value at the event horizon to the end state at $t_{\text{final}}=1000$.
In the right column of Fig. \ref{RNBH}, $\Delta\phi\equiv\phi\left(t_{\text{final}}\right)-\phi_{s}$
is plotted as a function of $r$, where $\phi\left(t_{\text{final}}\right)$
and $\phi_{s}$ represent the scalar field of the end state and the
static scalarized black hole, respectively. Fig. \ref{RNBH} demonstrates
that $\triangle\phi$ is approximately on the order of $10^{-6}$
or below. This implies two key points: First, the equilibrium state
can be accurately described by the static scalarized black hole with
the scalar field profile $\phi_{s}$ (shown in the insets). Second,
our numerical results achieve an accuracy of around $10^{-6}$.

\begin{figure}[t]
\begin{centering}
\includegraphics[scale=0.8]{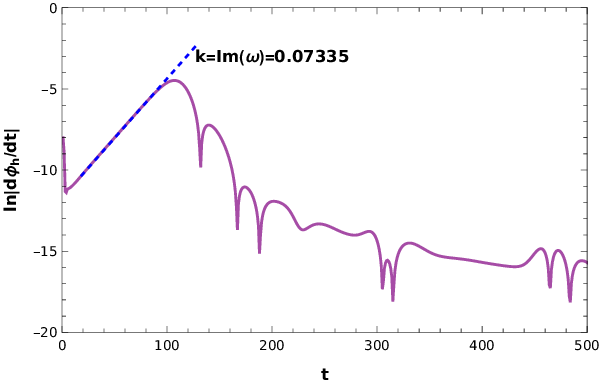} \includegraphics[scale=0.8]{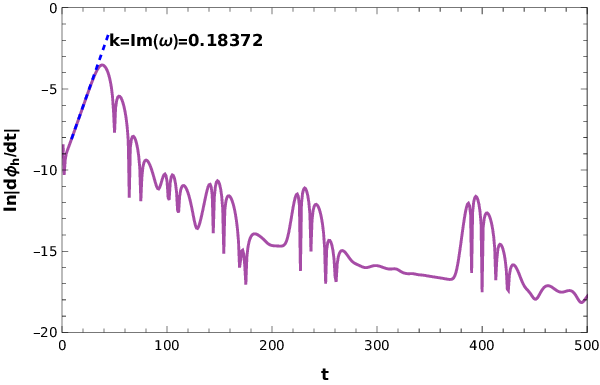} 
\par\end{centering}
\caption{Evolution of $\ln\left\vert d\phi_{h}/dt\right\vert $ during black
hole scalarization, for an initial RN black hole with $Q=0.9$ and
$M=1$. The left and right panels correspond to coupling constants
$\alpha=5$ and $\alpha=10$, respectively. The blue dashed line represents
the imaginary part of the unstable mode, matching the slope of $\ln\left\vert d\phi_{h}/dt\right\vert $
during the initial stages of scalarization. A larger $\alpha$ leads
to a higher imaginary part, indicating a stronger tachyonic instability.}
\label{RNlnphi}
\end{figure}

In Fig. \ref{RNlnphi}, the left and right panels depict $\ln\left\vert d\phi_{h}/dt\right\vert $
for black hole scalarization presented in the upper and lower rows
of Fig. \ref{RNBH}, respectively. The blue dashed line represents
the unstable tachyonic mode $\omega$, obtained by solving eqn. $\left(\ref{eq:eom_pert_omega}\right)$
with the boundary conditions $\left(\ref{eq:in-out bds}\right)$.
During the initial stages of scalarization, the slope of $\ln\left\vert d\phi_{h}/dt\right\vert $
closely matches the imaginary part of the unstable mode $\omega$.
This agreement validates our identification of the dominant instability
driving the scalarization process. Furthermore, a larger coupling
constant $\alpha$ leads to a more negative effective mass squared
in eqn. $\left(\ref{eq:delta phi}\right)$, resulting in a more pronounced
tachyonic instability. As shown in Fig. \ref{RNBH}, a higher $\alpha$
(right panel of Fig. \ref{RNlnphi}) corresponds to a larger imaginary
part of the unstable mode, accelerating the black hole's spontaneous
scalarization towards equilibrium.

\subsection{Nonlinear Stability of Scalarized Black Holes}

This section investigates the nonlinear stability of scalarized black
holes in the EMS model under spherical perturbations. As described
in Section \ref{subsec:Static-Black-Hole}, the effective potential
for photons in scalarized black holes can exhibit either a single
peak (corresponding to one unstable light ring) or a double peak (corresponding
to one stable and two unstable light rings). To analyze the black
hole's evolution under perturbations, we introduce the following quantities:
$\triangle\phi_{i}=\phi\left(t=0\right)-\phi_{s}$, $\triangle\phi_{f}=\phi\left(t=t_{\text{final}}\right)-\phi_{s}$
and $\triangle\phi_{fh}=\phi\left(t=t_{\text{final}}\right)-\phi\left(t=t_{\text{half-time}}\right)$.
Here, $\phi_{s}$ represents the scalar field of the initial static
scalarized black hole. Note that $\phi\left(t=0\right)$ is the initial
scalar field data incorporating the perturbation, and $\triangle\phi_{i}$
therefore quantifies the initial scalar perturbation. A small value
of $\triangle\phi_{fh}$ signifies that the black hole system reaches
a new equilibrium state after the perturbation. Meanwhile, $\triangle\phi_{f}$
reflects the deviation of the system's end state, $\phi\left(t=t_{\text{final}}\right)$,
from its initial scalarized black hole.

\subsubsection{Case without a Stable Light Ring}

\label{subsec:Light-Ring}

\begin{figure}[t]
\begin{centering}
\includegraphics[scale=0.77]{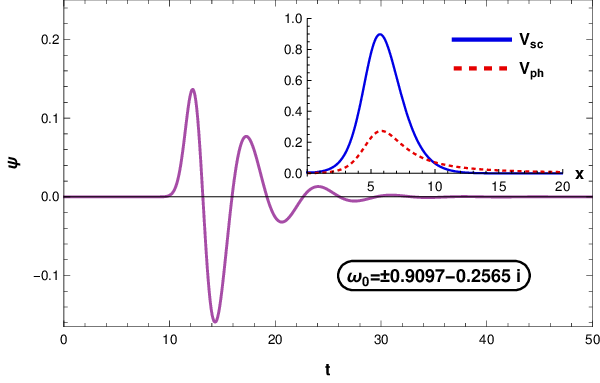} \includegraphics[scale=0.82]{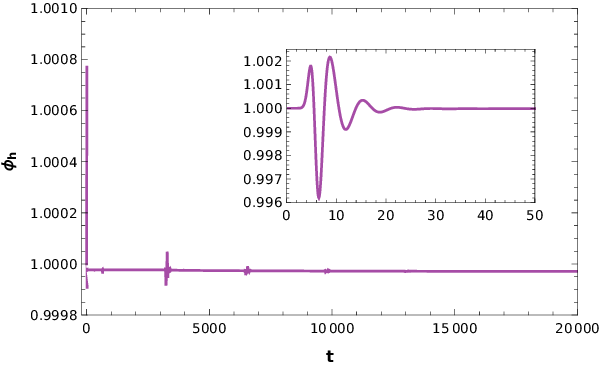} 
\par\end{centering}
\begin{centering}
\includegraphics[scale=0.77]{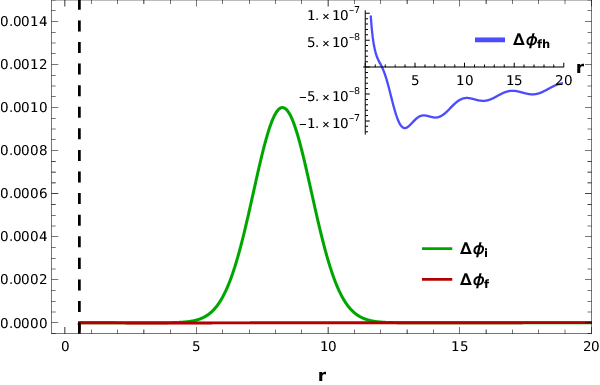} \includegraphics[scale=0.8]{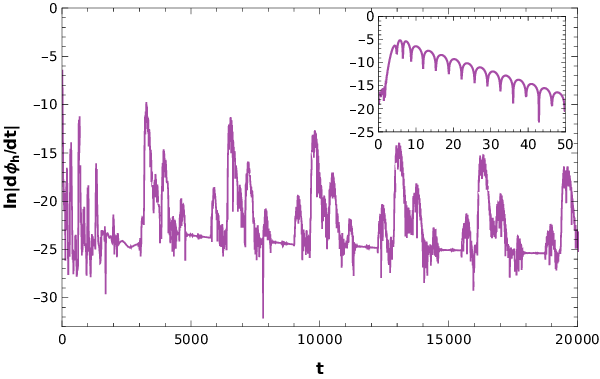} 
\par\end{centering}
\caption{Time evolution of a scalarized black hole with $Q=1.5418$ and $M=1$
for $\alpha=5$. \textbf{Upper-Left Panel}: The linear evolution of
the scalar perturbation damps out, indicating the linear stability
of the black hole. The inset shows the effective potentials for photons
and scalar perturbations, revealing the presence of a single unstable
light ring on the black hole's equatorial plane. \textbf{Upper-Right
Panel}: The nonlinear evolution of $\phi_{h}$ shows the black hole
system reaching equilibrium after initial oscillations. The inset
highlights the early evolution with a higher resolution, demonstrating
that the $\phi_{h}$ waveform closely resembles the linear case at
early times. \textbf{Lower-Left Panel}: The difference $\triangle\phi_{f}$
between the initial and final scalar fields, confirming long-term
nonlinear stability of the scalarized black hole. The inset depicts
the scalar field difference between the final state and the half-time
state $\triangle\phi_{fh}$, emphasizing the stability over the long
period. The horizon is represented by vertical black dashed lines.
\textbf{Lower-Right Panel: }The absence of growing modes in $\ln\left\vert d\phi_{h}/dt\right\vert $
signifies the absence of unstable modes in the nonlinear evolution.
The inset corresponds to the early evolution in the inset of the upper-right
panel.}
\label{alpha5}
\end{figure}

Fig. \ref{alpha5} examines a scalarized black hole with $Q=1.5428$,
$M=1$ and $\alpha=5$. The inset of the upper-left panel displays
its effective potentials for photons and scalar perturbations, revealing
the presence of only one unstable light ring on the equatorial plane.
The upper-left panel presents the linear evolution of a scalar perturbation
in this black hole, governed by eqn. $\left(\ref{eq:eom_pert_t}\right)$.
The dominant quasinormal mode $\omega_{0}$ is computed using eqn.
$\left(\ref{eq:eom_pert_omega}\right)$ and exhibits a negative imaginary
part. This indicates the linear stability of the black hole, as the
linear perturbation damps out towards both the event horizon and spatial
infinity.

The remaining three panels depict the fully nonlinear evolution of
the EMS system starting from the scalarized black hole, governed by
eqns. $\left(\ref{eq:metric eqs}\right)$ and $\left(\ref{eq:scalar eqs}\right)$.
The upper-right panel shows the long-term time evolution of $\phi_{h}$,
revealing that the scalar field evolves toward an equilibrium state.
The observed pulses result from partial reflections at the imposed
boundary at $x=1600$, as the radiative boundary condition cannot
fully dissipate the propagating fields. As anticipated, these pulses
exhibit a period of approximately $T\approx2\times1600=3200$, with
their amplitude diminishing significantly at later times. Additionally,
an inset plot presents the early evolution with four times finer temporal
resolution, ensuring high numerical precision albeit at increased
computational cost. Interestingly, the waveform of $\phi_{h}$ bears
resemblance to the linear case, suggesting that the black hole system
oscillates before settling into an equilibrium state.

To characterize the equilibrium state, we present $\triangle\phi_{i}=\phi\left(t=0\right)-\phi_{s}$
and $\triangle\phi_{f}=\phi\left(t=20000\right)-\phi_{s}$ in the
lower-left panel. Remarkably, the green line representing $\triangle\phi_{f}$
indicates that the final equilibrium state aligns with the initial
scalarized black hole. The inset shows $\triangle\phi_{fh}=\phi\left(t=20000\right)-\phi\left(t=10000\right)$
as a function of $r$, demonstrating the black hole's long-term nonlinear
stability. In the lower-right panel, we plot $\ln\left\vert d\phi_{h}/dt\right\vert $,
with an inset exhibiting the same quantity for the early evolution
at a finer resolution. Similar to the $\phi_{h}$ plot, this panel
shows periodic signal pulses arising from the radiative boundary condition.
Importantly, the plot of $\ln\left\vert d\phi_{h}/dt\right\vert $
reveals the absence of any unstable modes during the nonlinear evolution,
indicating the system's nonlinear stability.

\subsubsection{Case with a Stable Light Ring}

\begin{figure}[t]
\begin{centering}
\includegraphics[scale=0.77]{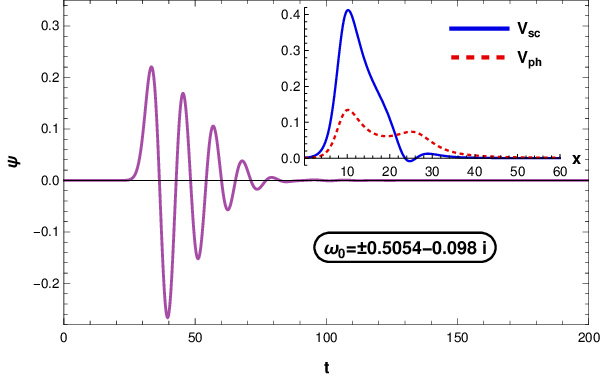} \includegraphics[scale=0.82]{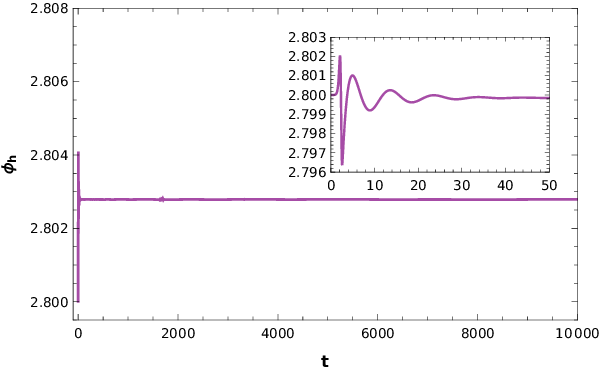} 
\par\end{centering}
\begin{centering}
\includegraphics[scale=0.77]{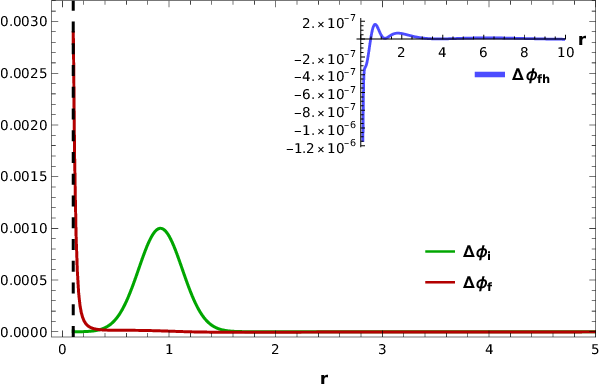} \includegraphics[scale=0.8]{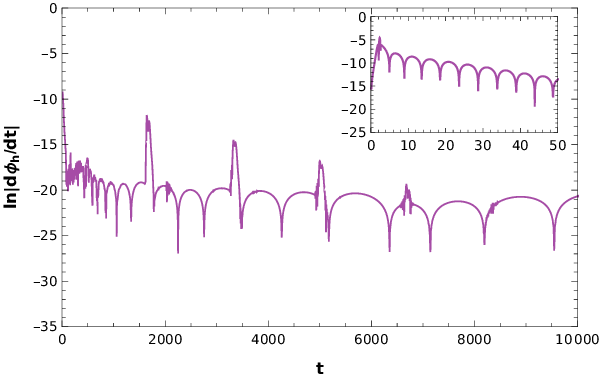} 
\par\end{centering}
\caption{Time evolution of a scalarized black hole with $Q=1.0516$, $M=1$
and $\alpha=0.8$, featuring a stable light ring. \textbf{Upper-Left
Panel}: The linear evolution of the scalar perturbation in the scalarized
black hole background damps out, signifying the linear stability of
the system. A dominant quasinormal mode $\omega_{0}$ with a negative
imaginary part further supports this stability. The inset shows the
effective potential for photons $V_{\text{ph}}$, where a local minimum
indicates the presence of a stable light ring. \textbf{Upper-Right
Panel}: The nonlinear evolution of $\phi_{h}$ demonstrates the black
hole's nonlinear stability against the light-ring instability. The
inset highlights the early evolution with a higher resolution, revealing
a close resemblance between the nonlinear and linear $\phi_{h}$ waveforms.
\textbf{Lower-Left Panel}: The scalar field difference between the
end state and the initial scalarized black hole, $\triangle\phi_{f}=\phi\left(t=10000\right)-\phi_{s}$,
confirms the long-term nonlinear stability of the scalarized black
hole. \textbf{Lower-Right Panel:} The waveform of $\ln\left\vert d\phi_{h}/dt\right\vert $
indicates the absence of unstable modes during the nonlinear evolution.
The inset corresponds to the early evolution depicted in the inset
of the upper-right panel. Periodic signal pulses observed are attributed
to numerical noises arising from the radiative boundary condition.}
\label{alpha08}
\end{figure}

In Fig. \ref{alpha08}, we initially place a scalar perturbation around
the stable light ring of the scalarized black hole with $Q=1.0516$,
$M=1$ and $\alpha=0.8$. The upper-left panel illustrates the linear
evolution of this perturbation within the background of the scalarized
black hole. Meanwhile, the remaining panels exhibit the fully nonlinear
evolution starting from the scalarized black hole. The inset of the
upper-left panel displays the effective potentials for photons and
scalar perturbations, revealing a stable light ring at the local minimum
of $V_{\text{ph}}$. Notably, the scalarized black hole exhibits a
dominant quasinormal mode $\omega_{0}$ with a negative imaginary
part, indicating its linear stability against the applied perturbation.
This is further corroborated by the damping of the perturbation observed
in the upper-left panel. Considering backreaction effects, the upper-right
panel presents the nonlinear evolution of the scalar field, indicating
the black hole's nonlinear stability. The inset highlights the early
evolution with a four times finer resolution, demonstrating its resemblance
to the linear case.

The lower-left panel shows $\triangle\phi_{fh}=\phi_{h}\left(t=10000\right)-\phi_{h}\left(t=5000\right)$,
a small value indicating that the system reaches equilibrium by the
end of the simulation. The end state closely resembles the initial
static scalarized black hole, as shown by the red line representing
$\triangle\phi_{f}$. However, numerical errors are evident near the
event horizon. The inset of the upper-right panel suggests that a
higher resolution simulation could mitigate these deviations. Furthermore,
the lower-right panel depicts $\ln\left\vert d\phi_{h}/dt\right\vert $,
with the inset corresponding to the early evolution in the upper-right
panel. This waveform of $\ln\left\vert d\phi_{h}/dt\right\vert $
signifies a lack of unstable modes in the nonlinear evolution. Similar
to Fig. \ref{alpha5}, the plot displays periodic signal pulses with
a period of $T\approx1600$, gradually diminishing in magnitude over
time.

\begin{figure}[t]
\begin{centering}
\includegraphics[scale=0.77]{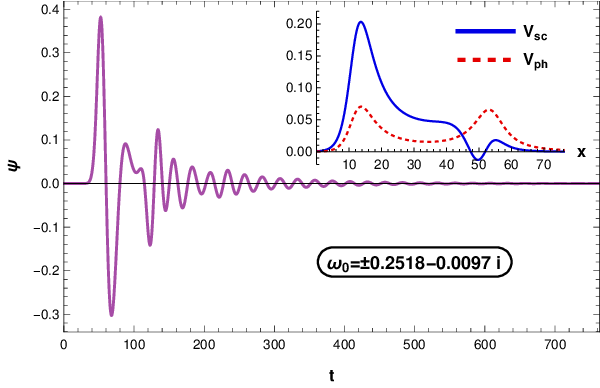} \includegraphics[scale=0.82]{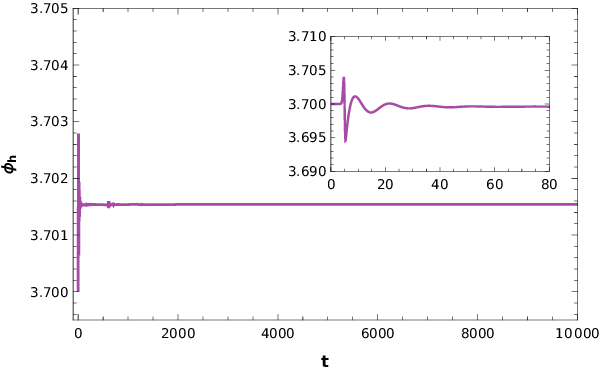} 
\par\end{centering}
\begin{centering}
\includegraphics[scale=0.77]{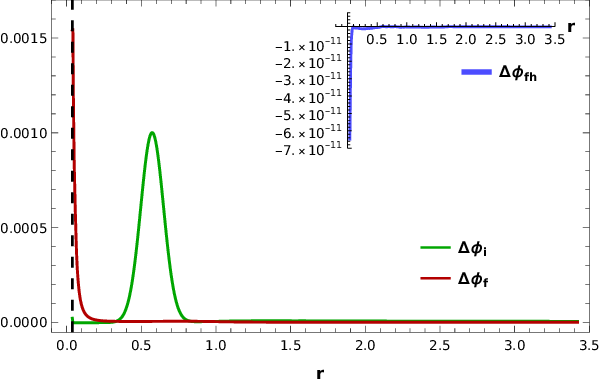} \includegraphics[scale=0.8]{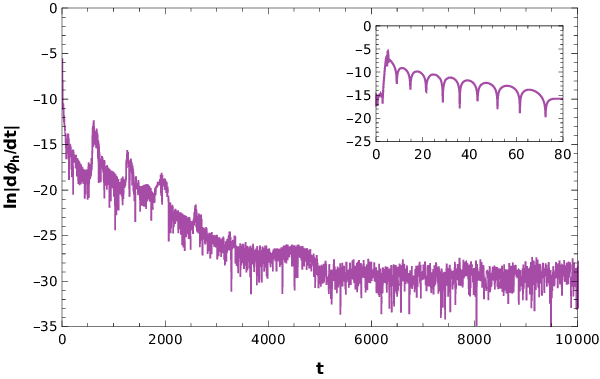} 
\par\end{centering}
\caption{Dynamics of a scalarized black hole with $Q=1.0192$, $M=1$ and $\alpha=0.6$,
exhibiting a stable light ring and a wider photon potential well.
The upper-left panel shows the damping of the linear perturbation,
signifying the linear stability of the black hole. The absence of
the light-ring instability in the nonlinear evolution is confirmed
by the dynamics in the upper-right and lower-right panels. The insets
in these panels depict the early stages of the nonlinear evolution
with a higher resolution, revealing close resemblance to the linear
evolution.}
\label{alpha06}
\end{figure}

In a previous study \cite{Cunha:2022gde}, it was shown that a wider
well in the photon effective potential near the stable light ring
can exacerbate the light-ring instability, leading to a faster detection
of this instability. To further investigate the nonlinear stability
of light rings in black hole spacetime, we analyze a scalarized black
hole with $Q=1.0192$, $M=1$ and $\alpha=0.6$, as presented in Fig.
\ref{alpha06}. The inset of the upper-left panel reveals a well-separated
double-peak structure in the photon effective potential, resembling
that of a wormhole. Interestingly, despite this feature, the numerical
results in Fig. \ref{alpha06} suggest that the stable light ring
remains nonlinearly stable against spherical perturbations.

\section{Conclusions}

\label{Sec:Conc}

This study explores the dynamical evolution of spherically symmetric
black holes within the EMS model, where a scalar field couples non-minimally
to the electromagnetic field through an exponential coupling function.
In the absence of a scalar field, tachyonic instabilities can trigger
spontaneous scalarization, which gives rise to scalarized black holes
emerging from RN black holes. Our numerical simulations confirm that
the evolution of RN black holes eventually stabilizes, leading to
the formation of static scalarized black holes. Interestingly, for
specific parameter ranges, the resulting scalarized black holes can
harbor a stable light ring on the equatorial plane.

While the existence of a stable light ring has been linked to potential
nonlinear instabilities that could eliminate the stable light ring,
our investigation did not yield evidence for such behavior. We studied
the long-term nonlinear evolution of scalarized black holes, both
with and without a stable light ring, under spherically symmetric
scalar perturbations. The simulations consistently showed that the
black holes undergo an initial oscillatory phase due to the perturbations.
However, they ultimately settle into an equilibrium state closely
resembling the initial configuration. This behavior signifies the
nonlinear stability of the scalarized black holes, indicating that
stable light rings in spherically symmetric black holes are resilient
against spherical perturbations.

Our focus here has been on spherically symmetric perturbations. However,
non-spherical perturbations with high angular frequencies are more
susceptible to becoming trapped around stable light rings \cite{Guo:2021enm},
consequently increasing their potential to trigger the light-ring
instability \cite{Keir:2014oka,Cunha:2022gde}. Therefore, future
investigations should explore the nonlinear stability of stable light
rings in scalarized black holes against non-spherical perturbations.
Furthermore, the long-lived modes trapped by stable light rings can
become unstable due to the ergoregion instability in rotating spacetimes
\cite{Cardoso:2014sna}. Additionally, while Schwarzschild-AdS black
holes are stable against spherical perturbations, the trapping mechanism
suggests potential dynamical instability for Kerr-AdS black holes
\cite{Holzegel:2011rk,Holzegel:2011uu}. Inspired by these findings,
it would be valuable to investigate the instability of scalarized
Kerr-Newman black holes, recently constructed in \cite{Guo:2023mda}.
\begin{acknowledgments}
We are grateful to Yiqian Chen and Shenkai Qiao for useful discussions
and valuable comments. This work is supported in part by NSFC (Grant
No. 12105191, 12275183, 12275184 and 11875196, 12347133, 12105126,
12247101). 
\end{acknowledgments}

 \bibliographystyle{unsrturl}
\bibliography{ref}

\end{document}